# Careers in astronomy in Germany and the UK

**We discuss the outcomes of surveys addressing the career situation of astronomers in Germany and the UK, finding social and cultural differences between communities as well as gender bias in both.**

In 2012, Fohlmeister and Helling published the results of the AstroFrauenNetwork survey of German women astronomers, asking about the job situation of women in Germany and of German women abroad. This first survey was carried out in 2011 and it had a response rate of more than 80% (61 participants), from women astronomers at all academic levels from doctoral students to professors as well as some who had left the field. In that first paper we surveyed the job situation and analysed different indicators for career development, reflecting on the impact of the personal situation of female astronomers in Germany: networking and human support were among the most important factors for career success, as well as work experience abroad and mentoring supervisors. We found that prejudice is abundant. We identified reasons why women are more likely to quit astronomy after they obtain their PhD degree than men and we gave recommendations for students on steps to take towards a successful career path in astronomy.

That first AstroFrauenNetzwerk (AFN) survey (Fohlmeister and Helling 2012) involved women only. Inspired by the feedback we received, we increased and diversified our sample to identify potential gender-specific differences. We performed the same survey with male astronomers working in Germany and German astronomers working abroad *before* the female-only results were published. To investigate if our results are representative for a larger astronomical community or if they were specific to German astronomers, the same survey was conducted with members of the Royal Astronomical Society (RAS). We report on the results of these surveys in this paper, including the response rates of the different communities.

Fohlmeister and Helling (2012) summarized the statistics on the gender distribution of the German Research Foundation (DFG), the Astronomical Society of Germany (Astronomische Gesellschaft, AG), the International Astronomical Union (IAU), the Max Planck Society (MPG) and the German Physics Society (DPG). These numbers have changed only marginally or have not been updated since the first paper was published.

## Where are women astronomers?

In the International Astronomical Union, Germany is represented by 506 members. Compared to 2010, the percentage of female German astronomers increased from 9% to 10%. A similar number of astronomers (527) from the UK hold membership in the IAU, but with a higher percentage of women (14%).
The percentage of women in the RAS is 16% (out of about 3500 members including astronomers and geophysicists; private communication RAS secretary). About 800 astronomers hold membership in the Astronomische Gesellschaft, with a smaller percentage of women: 11% (Schielicke 2013).
The percentage of female participation in physics programs and projects funded by the German Research Foundation (DFG) was 8% in 2010, corresponding to the average fraction of female physics professors in Germany (7.2%) (DFG 2012). The number of female directors among the nine research institutions of the Max Planck Society in Germany that carry out research in astronomy and astrophysics has changed: one directorship out of 32 is held by a women (see http://www.mpg.de/institutes). There is one female director at the two Leibnitz institutes in Germany that perform research in astrophysics.
On a more local scale, the organizing committee of the annual meeting of the Astronomische Gesellschaft (AG) in Hamburg in September 2012 evaluated their data with respect to female/male participants: 20% were women out of a total of 293 participants. 21% (5 in number) of the invited plenary talks (review and highlight talks) were given by women, and thus 79% by men. Of the plenary review talks, three were given by women and eight by men. Hence, 27% of the most prestigious talks were given by women. The fraction of female speakers in the total of 12 splinter meetings was 16% on aver- age, with a minimum of zero female speakers in one splinter and a maximum of 33% in the exoplanet splinter. The latter was organized by two female astronomers.

The number of female plenary review speakers at the 2013 AG meeting in Tübingen is one out of six, and six women and six men were invited highlight or plenary speakers. For the UK National Astronomy Meeting (NAM) 2012 (Manchester) and 2013 (St Andrews) the numbers were constant: two female invited plenary review speakers versus six male col- leagues, thus 25% of the most prestigious talks were given by women.

## Response rates

As in Fohlmeister and Helling (2012), the survey was conducted by collecting data via an online questionnaire, originally designed ahead of the annual meeting of the AFN during the annual AG meeting in Heidelberg in September 2011. Then we performed three complementary follow-up surveys, with exactly the same questions:

- Male astronomers working in Germany and German astronomers working abroad were approached via email in March 2012. For this, a list of 200 astronomers was compiled from astronomy institutes in Göttingen, Hamburg, Jena, Bonn, Würzburg and München as well as for German astronomers working abroad. 71 men responded – 35% of those approached. This survey was carried out *before* the results of the first survey were published.

- At the bilateral annual meeting of the RAS and the AG in Manchester in March 2012, both genders attending were asked to respond, with the support of the local organizers. 61 out of 934 conference attendees participated in the survey. The response rate of 6% was too low for the results to have significance and *they are not considered further here*.

- With the help of the Committee for Women in Astronomy and Geophysics (CWIAG) of the RAS, RAS Points of Contact in UK universities and research institutes were contacted by the RAS in June 2012: 277 RAS members (108 male, 169 female) participated out of about 3000 RAS members that could have received and read the invitation. This is a response rate of ~9%. RAS membership is open to astronomers and geophysicists, including journalists, teachers and amateurs, but because the invitation addressed astronomers, we expect the response rate among professional astronomers to be highest, and higher than this figure indicates.

We used the following sets of survey data: German women (from the autumn 2011 survey, Fohlmeister and Helling 2012) with a response rate of ~80% (61 out of 73 women) and German men (from the March 2012 survey above), with a response rate of ~35% (71 out of 200). Both samples are of comparable size (regarding the absolute number of participants), but the results have to be viewed with the dissimilar response rates in mind. This combined German sample is compared to the survey results from the UK/ RAS survey, a sample that provides by far the best statistics with respect to the absolute number of participants.

Our data define a representative sample of the astronomical community working in Germany and in the UK. It includes female and male astronomers from all academic levels, as well as astronomers who have left the field and work outside astronomy (see figure 1). It is also reasonable to expect that the response rate was generally highest for professional astronomers as they are easier to reach, and also because this survey is more relevant to them.

## Positions

Figure 1 shows the distribution of the survey participants over the positions held at the time of the individual surveys. In both samples, post- docs form the largest pool of participants. The percentage of female postdocs was higher than that of male postdocs in both countries at the time the surveys were taken. More female PhD students than male postdocs took part in the UK/RAS survey. The PhD students were the second largest pool of astronomers in Germany in the survey. Male professors/lecturers/tenure-track astronomers make the second largest pool in the UK/RAS sample.

The relative distribution of professor/lecturer/tenure-track positions between women and men is comparable in both countries: 16% male vs 7% female in Germany and 38% male vs 18% female in the UK. The position of junior research-group leader does not exist as such in the UK, hence the low numbers. In the German sample, more female than male astronomers appear to be junior group leaders. It should be interesting to see how many young astronomers can secure a long-term career in astronomy and which individuals will achieve this.

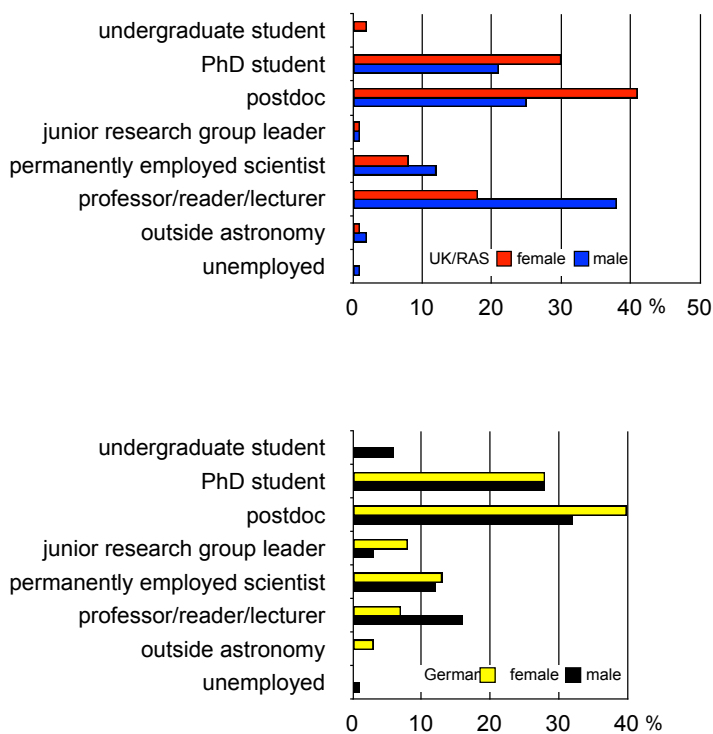

**1: Academic positions held by participants in the UK/RAS sample (top) and in the German sample (bottom). Note that the end dates differ somewhat due to the different times when the surveys were conducted: German women, August 2011; German men, March 2012; UK/RAS, March 2012.**

## Ages

The age distribution of the participants peaks around 31–35 years for both genders in the German sample (figure 2, bottom). The age distribution is shifted to younger ages for women in the UK/RAS sample, peaking in the age group 26–30 (figure 2, top). The fractions of male astronomers are almost the same in the age groups 26–30 and 31–35 in the UK/RAS sample, while the number of female astronomers is always larger in the 21–25, 26–30 and 31–35 yrs groups. The number of female astronomers responding is almost twice as high as the number of men in the 26–30yrs group. This reverses in the 41–45yrs group: the number of female astronomers is only a third of that of the male astronomers. Only male astronomers appear in the 56–60yrs and 61–65yrs groups. The German sample (figure 2, bottom) is more homogeneous in the age groups 21–25, 26–30 and 31–35. A larger fraction of women populate the 36–40yrs group, while men and women have similar numbers from 41–50yrs. Male astronomers dominate the 51–55 group. Only male astronomers appear in the 56–60 and 61–65yrs groups in the German sample, too. This suggests that we did not reach any of the very few German female astronomers in these age groups. We note the coincidence between the two independent samples of two different European countries.

The age grouping is biased to those groups of participants for whom this survey matters most. We refer here to the response rates outlined earlier: the survey of German male astronomers had a response rate of 35% compared to 80% in the German female astronomer survey. We do not have similarly detailed data for the UK/RAS sample.
We refer the reader to demographic and gender statistics as, for example, published in the *She Figures* report of the European Commission (2012). For example, *She Figures* demonstrates that more men than women obtained a PhD in science, mathematics and computing for their 2010 survey in UK and in Germany, and that more men than women work in natural science research in the higher education sector.
There are both more older male scientists than female scientists of similar age, and more men in higher positions in both the German and UK/ RAS sample. Both findings are not new to the respective communities.

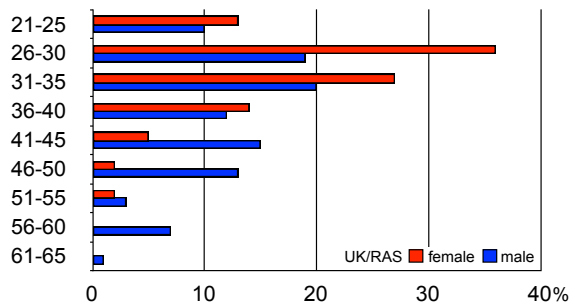

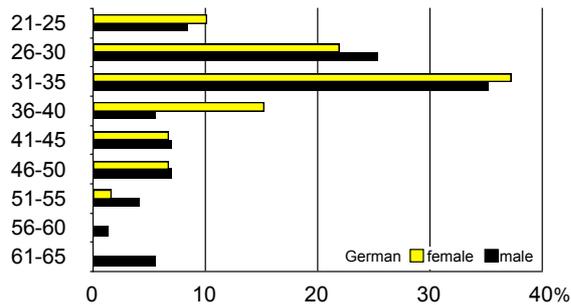

2: Age distributions of the survey participants in the UK/RAS sample (top) and in the German sample (bottom).

# Geography

We investigated the country of origin and place of work of the survey participants (figure 3). The fraction of female astronomers in the UK/ RAS sample who grew up in the UK, in another country in Europe or outside Europe/overseas is one-third each, whereas among the male astronomers of the UK/RAS sample about 50% grew up in the UK. The UK/ RAS sample shows a more international distribution than the German sample, where about 70% of both genders grew up in Germany, indicating the underlying astronomical community being less international.

A far higher fraction of men (93% in the German sample, 71% in the UK/RAS sample) than women (73% in the German sample, 48% in the UK/RAS sample) currently work in Germany/ the UK. Out of those who grew up in Germany/ UK, more men (93% in the German and in the UK/RAS sample) than women (76% in the German sample, 88% in the RAS sample) currently hold a position in the country of their origin. Our survey does not distinguish between those astronomers who returned to their home country after a stay abroad and those who remained in their home countries.

All survey participants were asked to comment on whether they would like to return to their home countries. While some female German astronomers expressed their desire to stay outside Germany, the men in the German sample who answered this question (31%) would prefer to live in the country of their origin, because they like to be close to their families and friends or raise their kids in their home country/culture.
Some male participants state that they are simply not willing to work abroad and prefer the working conditions in Germany. In the UK/ RAS sample, those who answered the question (47% of the female and 31% of the male participants) state that they would prefer to work there ("home is home and being home is something I guess everyone would prefer") but often do not because of worse working conditions in their respective home countries or the two-body problem (i.e. a partner with a job elsewhere).

**3: Country of origin and current workplace for the UK/RAS sample (left) and the German sample (right).**

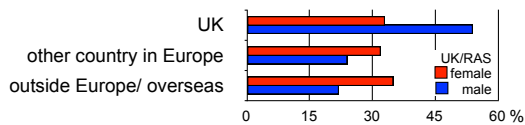
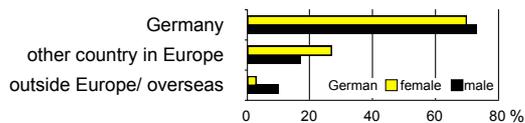
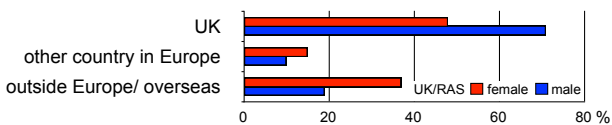
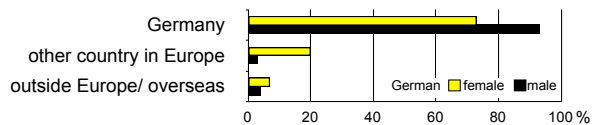

# Length of contracts

The duration of contracts for early-career astronomers is very heterogenous (figure 4). Short-term contracts (up to two years) are more common in the German than in the UK/RAS sample, but three years is the most common length for both. With the UK/RAS sample having a high fraction of professor/reader/lecturer positions, the fraction of survey participants having a permanent contract is highest among male UK/RAS astronomers (45%), comparable for German male and UK/RAS female astronomers (about 20%) and lowest for female German astronomers (14%). Female German astronomers represent the highest fraction of longer contracts (with a duration of up to five or more than five years), typical of positions that aim to qualify for permanent positions.

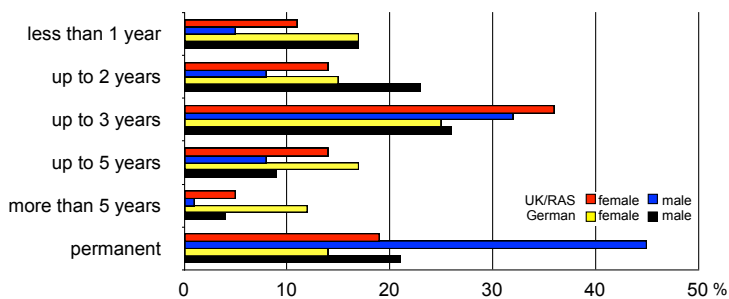

**4: Duration of current contract.**

# Finding a job

Figure 5 shows the different approaches to finding a position, ranking the most successful ones for all survey groups first. Personal contacts are most important for German astronomers. Applying for an openly advertised position on the internet is ranked highest for female and male UK/RAS astronomers and second highest for female German astronomers.

Male German astronomers more often change position within their institution, are personally asked to apply for a job or are recommended by a mentor; in the UK/RAS sample, more men than women are asked to apply or move within an institute. The fraction of astronomers recommended by a mentor is highest for women in the UK/RAS sample (28%; see also figure 6).

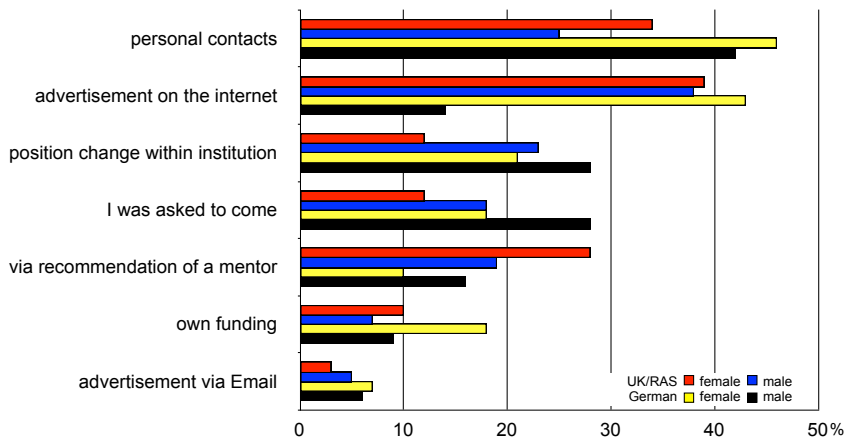

**5: How did you find your job?**

We also asked about the most important criteria on which the survey participants based their choice of a job (figure 7): working on an interesting project is most important for all, followed by the reputation of an institute and working conditions. Having family nearby and working on an interesting project is most important for German astronomers. A change of institute, promotion prospects and salary are the least important for all groups.

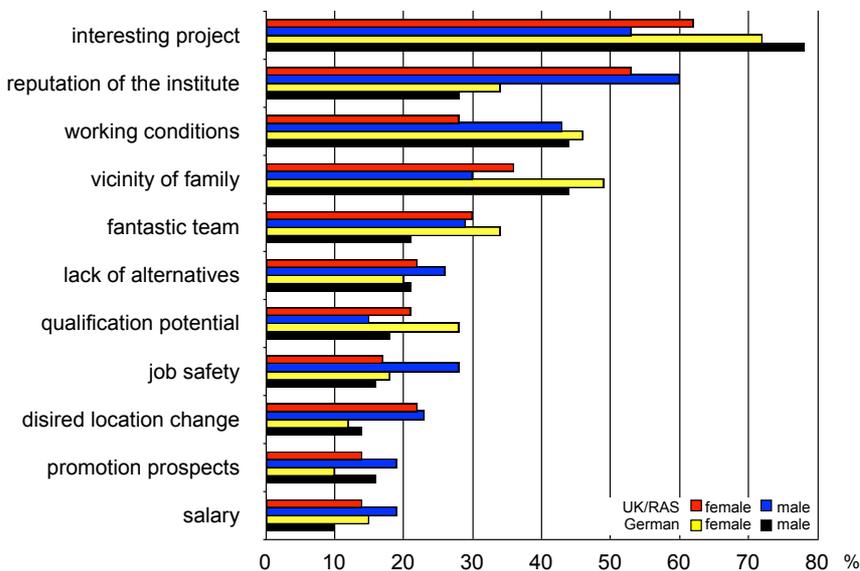

**7: Why did you choose your current position? Respondents in percent who selected the categories shown (multiple answers possible). The plot sorts the responses by frequency: "interesting project" was most often chosen.**

## Job satisfaction

We asked about job satisfaction and how content survey participants were with different aspects of their job, on work and personal criteria. The results in figure 8 show that overall satisfaction is good. Both sample groups and both genders are on average most happy with their scientific projects, working conditions, workplace equipment and opportunities to develop their own ideas. Groups evaluate the travel-funding situation differently: UK/RAS astronomers are less happy with their travel funds than the German sample. All four groups are less content with their promotion prospects, although this criterion was not judged to be important by any group when choosing a new job (figure 7). For job security, UK/RAS men are happiest and German women least happy, reflecting the underlying job situation: 45% of the UK/RAS male participants and only 14% of the German female astronomers hold a permanent position (figure 4).

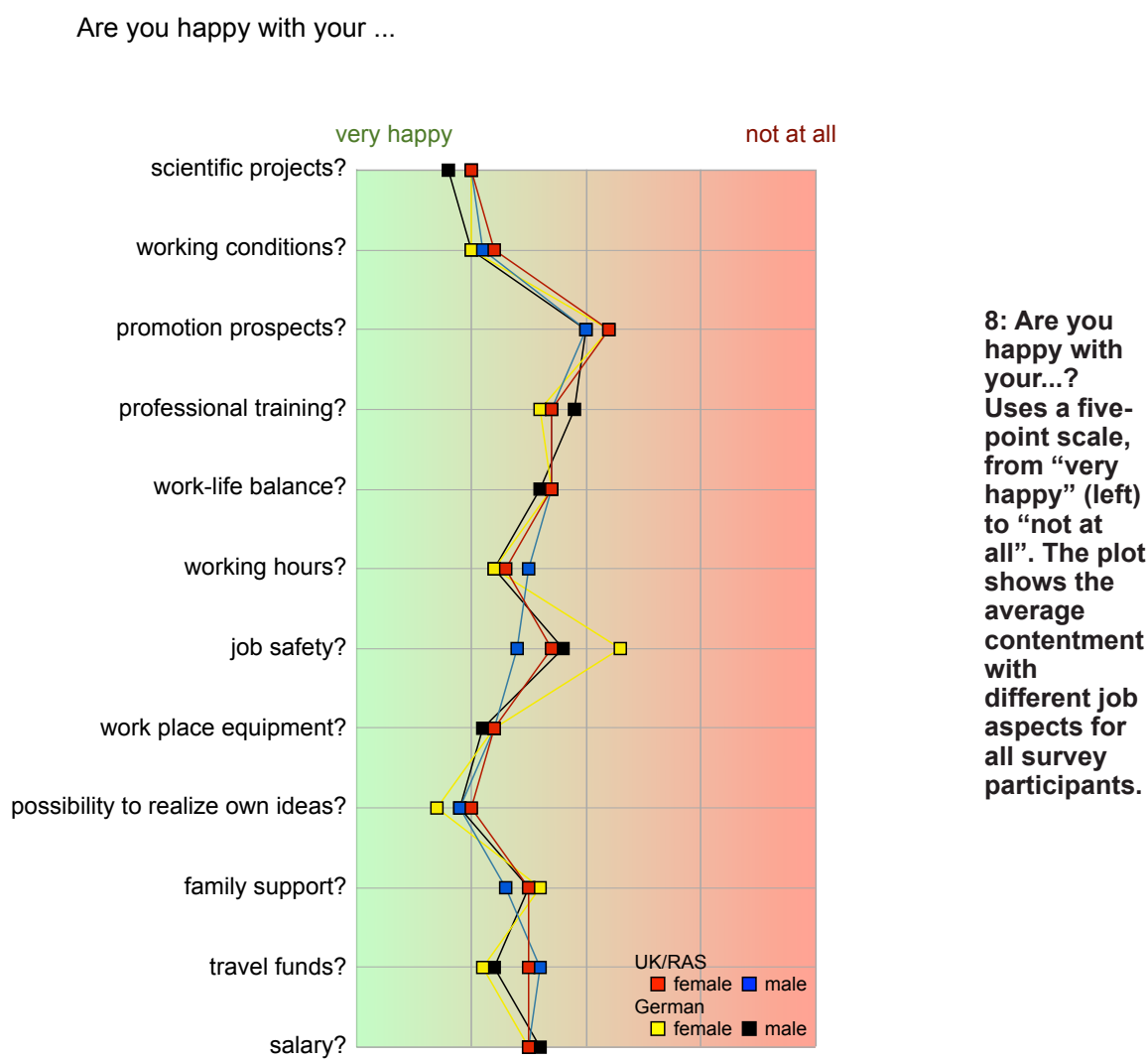

8: Are you happy with your...? Uses a five-point scale, from "very happy" (left) to "not at all". The plot shows the average contentment with different job aspects for all survey participants.

Participants were asked if they wanted a long-term career in astronomical research (figure 9). Male German astronomers and all astronomers from the UK/RAS show a similar distribution; German female astronomers seem less sure. This is reflected in the numbers collected during the surveys: more than 50% of the German male and UK/RAS astronomers state that they absolutely want to continue research in astronomy in the long term, the majority of the German women (55%) would only do it "if possible". Only 3% of the German women indicate that they "don't need to". This answer was more common among the other three groups of participants.

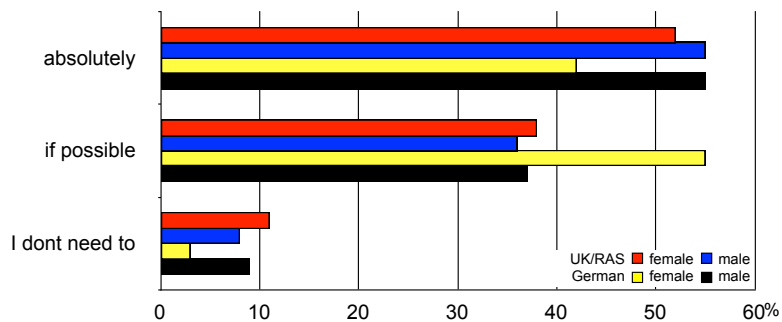

**9: Do you want to perform astronomical research in the long term?**

## Private life

More than 70% of all survey participants have a partner or spouse. The fraction of women with an astronomer or physicist partner is much higher than that for men in both the German and UK/RAS sample (figure 10). Men more often find their partner outside science. UK/ RAS sample members more often have partners but, among them, fewer live in the same city.

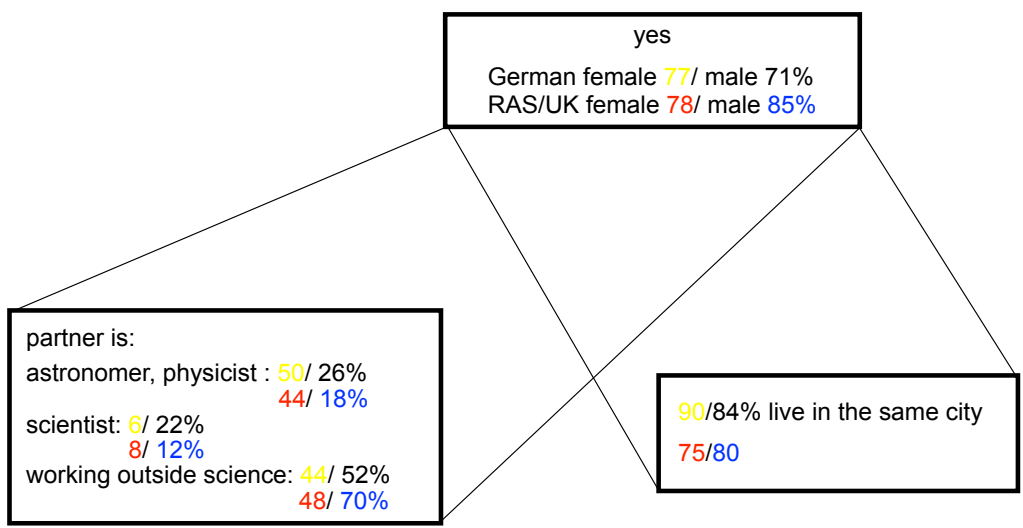

**10: Do you have a partner/ spouse? She/he is...? Partner lives in the same city?**

About one-third of German and female UK/ RAS astronomers who responded have children (figure 11). The highest fraction of astronomers with children is in the sample of male UK/RAS members (44%). This is the group that also has the highest fraction of people who don't want to have children (41%). This indicates that male UK/RAS members are more likely to have children if they want to have them, independent of their job situation (figure 12). Almost one-third of the German male and UK/RAS female participants state that they do not want to have children. Most German female survey participants (89%) would like to have at least one child.

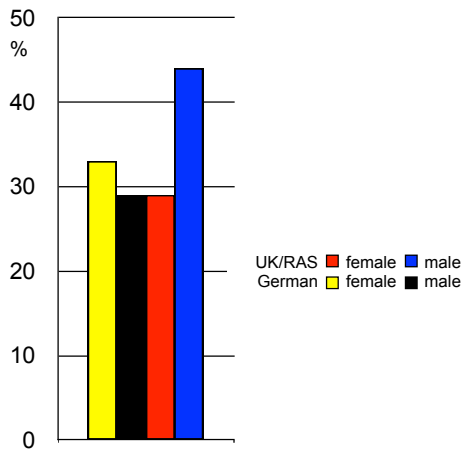

**11: Percentages of participants with children.**

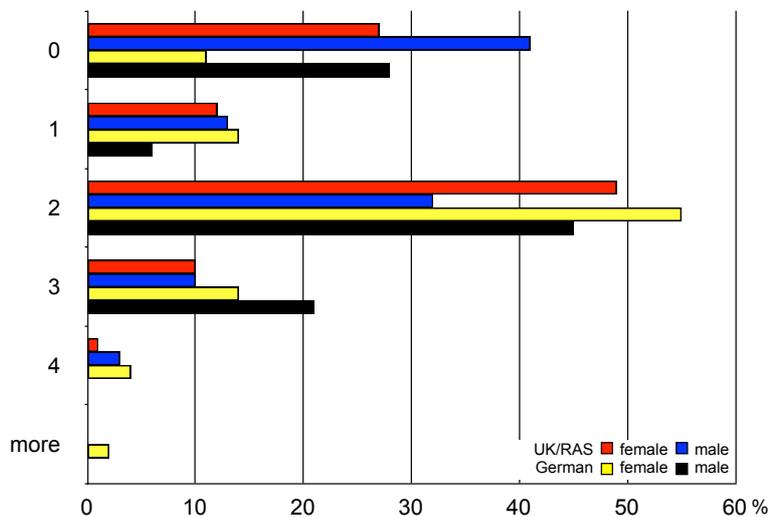

**12: How many children would you like to have?**

Figure 13 shows how the survey participants who have children see the resulting changes to their professional role. In the German female sample, women with children feel very restricted in mobility. Women from both samples pointed out that it is harder to combine job and family when they have children. However, women with children also feel generally more balanced and motivated. UK/RAS women also feel more restricted in their mobility than men. Otherwise, the differences are less emphatic in the UK/RAS sample than in the German sample. The percentage who are parents and who have problems concentrating arising from the family workload (>40%), and those who feel more balanced and motivated (>20%) is almost the same among German women and UK/RAS men and women. Problems keeping up after having children affect half UK/RAS members but less than 15% of the German sample. Over all participants, German men most appreciate the positive aspects of having children ("I feel more balanced and motivated" and "I am as active as before") and the negative aspects seem much less important compared to the female groups in our samples.

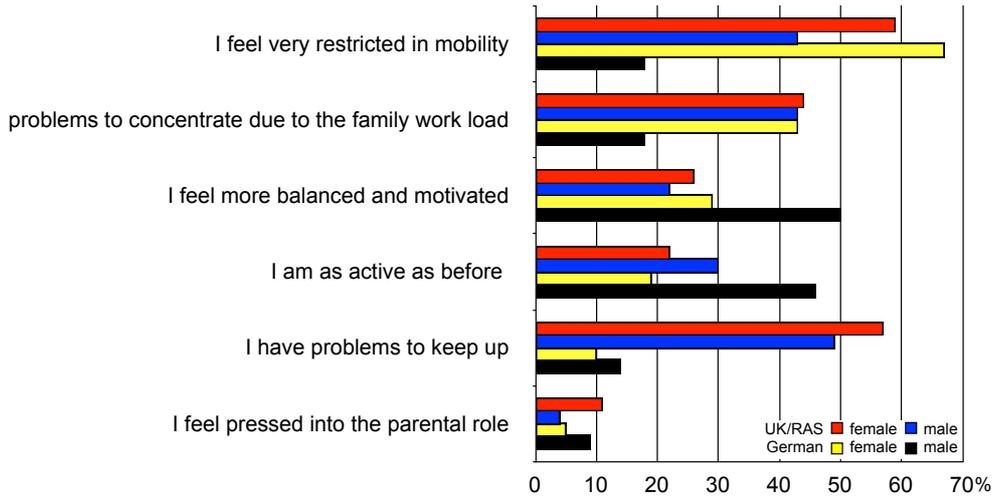

**13: How did your professional role change due to children?**

## Career development

In our first survey of female German astronomers we asked how many **professional trips abroad** (longer than three months) they had made. We found that despite the positive career impact of working abroad, 29% of the female astronomers had never done so. The 71% of German women who have worked abroad compares to less than half of the German men and the UK/RAS men and women who have done so (figure 14).

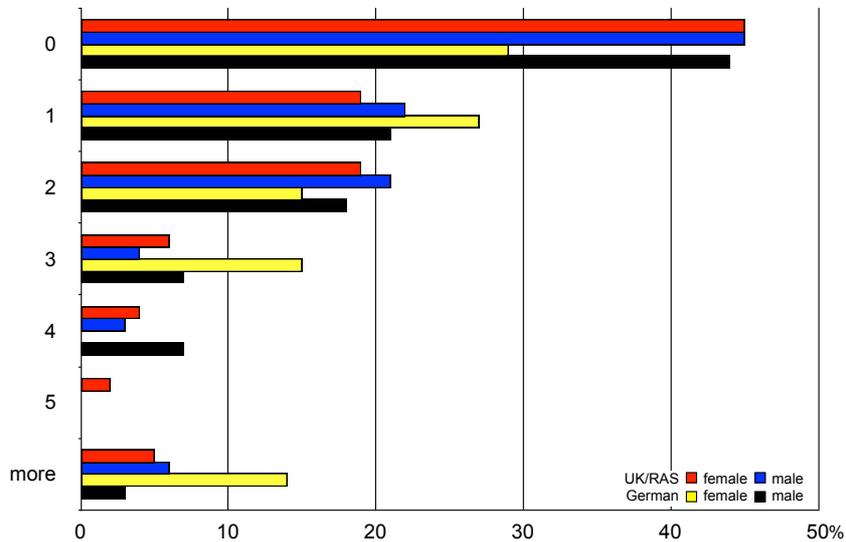

**14: How many stays abroad (for more than three months) did you complete?**

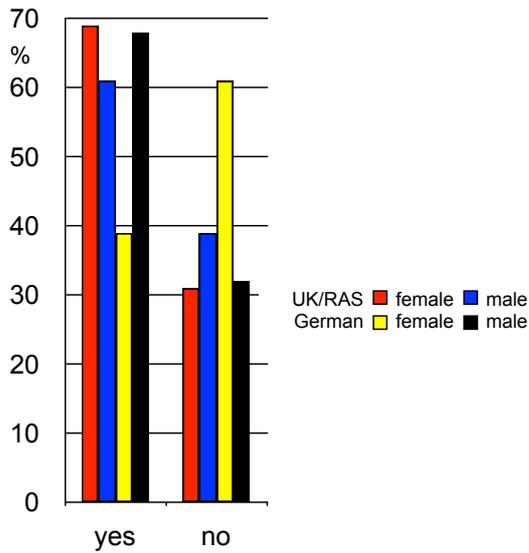

The survey also asked about the **importance of mentors**. The results (figure 6) show that, out of all participants, only 39% of the female astronomers in the German sample, but more than 60% of the other three groups have been supported by a mentor and felt that it was very helpful for their career.

**6: Did/do you have a mentor who helped you to advance?**

When asked what the **most important factors for a career in astronomy** are, national and gender differences appear more pronounced (figure 15). The top three factors identified by all groups are the "number of refereed papers", the "quality of research" and "networking". The second most important criteria were weighted to be "work on a hot topic" and "strong mentor". Criteria that were judged as more important by women than men are "regular presence at conferences", "flexibility", "support by family" and "organizing and negotiation skills". The "quality of their own research", "creativity" and "ability to work in a team" is judged more important by UK/RAS astronomers than by German ones.

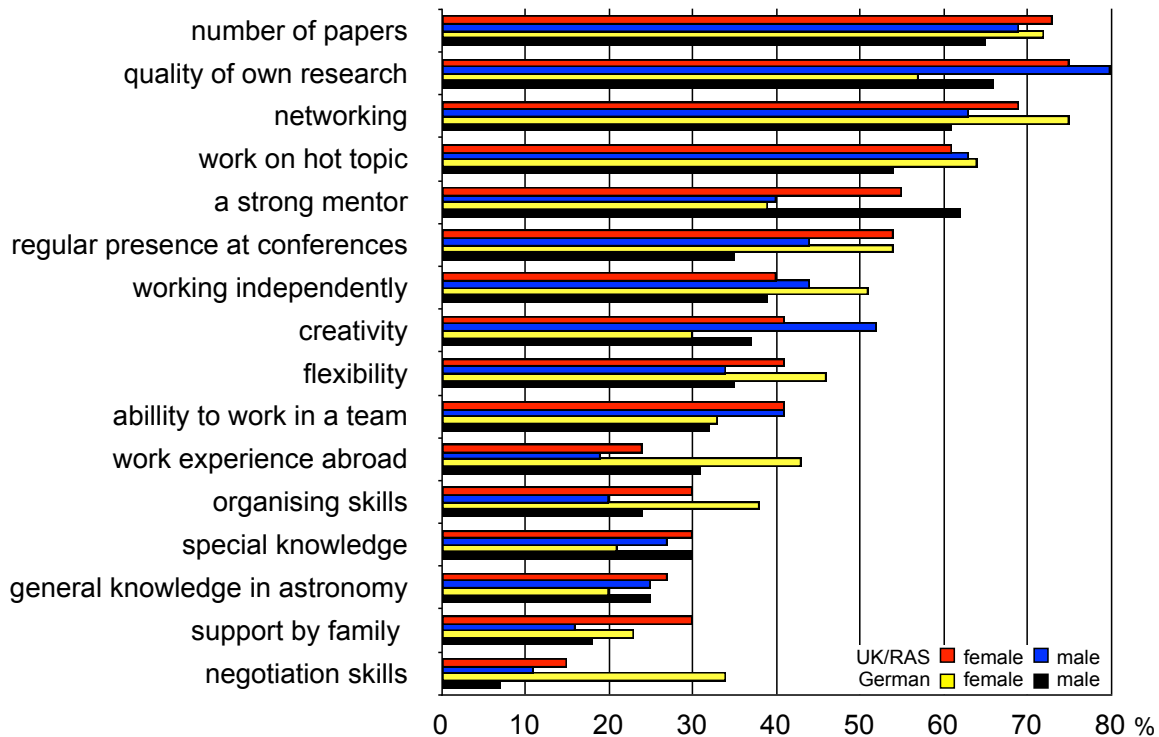

**15: What are the most important factors for a career in astronomy? Percentage who selected factors (multiple answers were possible), sorted by frequency of mention.**

The questionnaire allowed participants to describe what they had found **helpful or harmful for their career**. The overall answers are very similar to the ones obtained from German women astronomers in 2011 (Fohlmeister and Helling 2012). They identified as helpful: having good mentors/advisors, early career success from prestigious PhD programs or postdoc projects, working in world-leading institutes, good collaborations, hard work and well defined projects, going abroad, attending many international conferences, networking, and persistence even when the future looked grim.

However, the wording of answers to the question "What was helpful or harmful for the career?" in the survey was utterly different between men and women. Women tended to use descriptive wording of helpful career circumstances, such as having had a strong mentor or good advice. Men were much less shy about praising themselves, using statements such as "I chose the right people and projects", "I successfully write proposals", "I have original ideas" and "I have expertise that could not be replaced". Far more German men than UK/RAS men made these types of statements. We should also note that although this sample might be statistically representative, the answers quoted above may do injustice to individuals of both genders. Male German and UK/RAS astronomers also gave simply "luck" as a reason for their success in astronomy, an expression that did not appear in the answers of female astronomers.

When describing what was harmful for their career, the answers by all groups are mainly the contrary of the items listed above. German men also complain about "hierarchical structures", "egoistic" supervisors or "arrogant" collaborators – words that do not appear in the answers of the female German and UK/RAS participants. But female and male survey participants from both communities report lack of job security, contracts that are too short for advancement and lack of mentors or visits abroad as harmful for their career. Male survey participants also list personal difficulties affecting their career, such as feeling homesick, having depression, problems with partners or difficulties balancing job and family. All four groups report that long-distance relationships or being tied to a partner and having children are harmful for their career. Female UK/RAS astronomers report that sexual harassment by senior professors or inappropriate comments from colleagues made them feel not good enough for a scientific career. Similar statements were not made in the survey of German female astronomers.

## Discrimination

In Fohlmeister and Helling (2012) we summarized situations and discriminating comments described by German female astronomers. In the male German and UK/RAS sample, participants were also asked if they had experience of workplace bullying, unfair treatment or comments and if those were gender-specific.

Among female UK/RAS astronomers, 95 (56%) noted their experience. Out of them, 25 state that they have never experienced discrimination or gender-specific bullying. The other 70 participants describe situations ranging from unfair comments to sexual harassment. Many of the women report that colleagues said they were shortlisted for a position or awarded a fellowship only because they are women ("quota women"), implying a lack of scientific potential. "Inappropriate comments", "unfair treatment", "a general men's club atmosphere" or "not taken seriously", "being looked down on" and "seen as being less able", "being ignored", "feeling isolated", "treated differently for being female" are among the experiences reported. Female scientists also point out that they have the impression that their "results are questioned more than those of male colleagues" and some report that they were considered less productive. Female astronomers also report that they were excluded from collaborations because of possible or existing motherhood, for going on maternity leave or working part time. The statements include that the above causes a lot of frustration and that, despite a strong interest in astronomical research, "it just isn't worth it".

In the German and UK/RAS male samples only a few survey participants say they remember any bullying and if so that this was most of the times not (evidently) gender-related but due to hierarchical structures and hard competition. Nevertheless, some men report situations where female colleagues were treated unfairly and finally left. Many male participants also recounted their impression that women are preferred for positions or fellowships just because the institute had to hire women, and that this "makes things doubly frustrating when having to apply on a tight job market".

# Summary

Inspired by the community and its feedback on our first paper on "Career situation of female astronomers in Germany", we set out to increase and diversify our survey sample in order to identify potential gender-specific and cultural differences. While the overall gender differences in the astronomy communities are largely similar (but differ in details) the surveys reveal different perceptions and awareness between them:

- More men than women work in senior positions, in both the German and the UK/RAS sample. This is reflected in the age and positional distribution of the survey participants and not new to the community. The UK/RAS sample is more international than the German sample.

- Most German women (>70%) spent at least one period of their professional life in astronomy abroad, whereas 45% of the German male astronomers and the UK/RAS sample have never worked abroad.

- Personal contacts play a more important role in finding a position for German than for UK/ RAS astronomers.

- More men than women are invited to apply for a position or to change roles within an institution in both samples.

- More men than women hold a position in the country of their choice.

- Women more than men find their job via application to an openly advertised position on the internet or by raising their own funding.

- German women are the group least often supported by mentors. All who have been supported found this very helpful for their career.

- German women feel the most restricted in their mobility due to children and having to cope with the family workload. The positive side of having children is most appreciated by German men, who feel the negative side much less strongly than other surveyed groups.

- All groups were contented, overall, in astronomy and most would like to work in astronomy in the long term.

- All participants point out that the field is competitive and that a career in astronomy is hard to balance with a private life in terms of living where you choose, having a partner or raising children. Women additionally have to cope with the gender bias.

*Janine Fohlmeister, Astronomisches Rechen-Institut, Zentrum für Astronomie der Universität Heidelberg, Germany.*
*Christiane Helling, SUPA, School of Physics & Astronomy, University of St Andrews, UK.*

*Acknowledgments*
*We would like to thank the local organizers of the NAM 2012 in Manchester, Robert Massey from the RAS, the Committee on Women in Astronomy and Geophysics (CWIAG), and the RAS for their support. Iris Traulsen, Inga Kamp, Stefanie Komossa, Eva Grebel and Jessica Agarwal are thanked for valuable comments, Rebekka Weinel for technical support and everyone who filled in the questionnaire.*
*A special thanks goes to the readers of the first paper and to those who provided feedback.*